\documentclass[overric,runningheads]{llncs}
\makeindex          % be prepared for an author index
\usepackage{graphicx}
\usepackage{hyperref}

\usepackage{subcaption}
\usepackage{wrapfig}
\usepackage{colortbl}
\usepackage{csquotes}
\usepackage[american]{babel}
\usepackage{booktabs}
\usepackage[utf8]{inputenc}
\usepackage{soul}
\usepackage{algorithm}
\usepackage[noend]{algpseudocode}
\makeatletter
\def\BState{\State\hskip-\ALG@thistlm}
\makeatother

\usepackage{multirow}
\usepackage{wrapfig,lipsum,booktabs}

\usepackage{amsfonts}
\usepackage{amsmath}
\usepackage{amssymb}
\usepackage{booktabs}
\usepackage{caption}
\usepackage{cite}
\usepackage{comment}
\usepackage{float}
\usepackage[utf8]{inputenc}

\usepackage{adjustbox}
\usepackage{todonotes}
\usepackage{wrapfig}
\usepackage{rotating}

\def\post#1{\ensuremath{{#1}\kern-.05ex\bullet}\,}
\usepackage{xcolor}

\DeclareMathOperator*{\argmin}{arg\ min}
\newcommand{\alignment}{\ensuremath{\gamma}}
\newcommand{\approxError}{\epsilon}
\newcommand{\costOfAlignmentFunc}{\ensuremath{\kappa}}
\newcommand{\costOfMove}{\ensuremath{c}}
\newcommand{\editDistance}{\ensuremath{\delta}}
\newcommand{\editDistanceSet}{\ensuremath{\Delta}}

\newcommand{\eventLog}{\ensuremath{L}}
\newcommand{\lang}{\ensuremath{\mathcal{L}}}
\newcommand{\lobo}{\ensuremath{\bot}}
\newcommand{\model}{\ensuremath{M}}
\newcommand{\modelSubset}{\ensuremath{M'}}
\newcommand{\modelTrace}{\ensuremath{\varphi}}
\newcommand{\naturals}{\ensuremath{\mathbb{N}}}

\newcommand{\optimalAlignmentCost}{\ensuremath{z}}

\newcommand{\process}{\ensuremath{P}}
\newcommand{\proxySet}{\ensuremath{\Omega}}
\newcommand{\reals}{\mathbb{R}}
\newcommand{\sequence}{\ensuremath{\sigma}}
\newcommand{\ubo}{\ensuremath{\top}}
\newcommand{\univAct}{\ensuremath{\Sigma}}
\newcommand{\univAlignments}{\ensuremath{\Gamma}}
\newcommand{\univMSet}{\ensuremath{\mathcal{B}}}

\usepackage{tikz}
\usetikzlibrary{arrows,backgrounds,calc,decorations.pathmorphing,decorations.pathreplacing,fit,matrix,patterns,petri,positioning, shapes}

\renewcommand*\abstractname{abstract}

\newcolumntype{M}{>{\begin{varwidth}{11.1cm}}l<{\end{varwidth}}} %M is for Maximal column

\usepackage{diagbox}
\begin{document}
	\mainmatter              % start of the contribution

	\title{Model Independent Error Bound Estimation for Conformance Checking Approximation}
	\titlerunning{Model Independent Error Bound Estimation}  % abbreviated title (for running head)

\author{Mohammadreza Fani Sani\inst{1} \and Martin Kabierski \inst{2}\and Sebastiaan J. van Zelst\inst{3,1} \and Wil M.P. van der Aalst\inst{1,3}}

\authorrunning{Mohammadreza Fani Sani et al.}   % abbreviated author list (for running head)
\tocauthor{Mohammadreza Fani Sani, Martin Kabierski, Sebastiaan J. van Zelst, Wil M.P. van der Aalst}
\institute{$^1$Process and Data Science Chair, 
	RWTH Aachen University, Aachen, Germany\\
	$^2$Department of Computer Science, Humboldt-Universität zu Berlin, Berlin, Germany\\
	$^3$Fraunhofer FIT, Birlinghoven Castle, Sankt Augustin, Germany \\	
	\email{\{fanisani,s.j.v.zelst,wvdaalst\}@pads.rwth-aachen.de, martin.bauer@hu-berlin.de} }

%\and 
 
% \email{{sebastiaan.van.zelst,wil.van.der.aalst}@fit.fraunhofer.de}\\
	
	\maketitle          % typeset the title of the contribution   
   	
	\begin{abstract}
	Conformance checking techniques allow us to quantify the correspondence of a process's execution, captured in event data, w.r.t., a reference process model.
	In this context, alignments have proven to be useful for calculating conformance statistics.
	However, for extensive event data and complex process models, the computation time of alignments is considerably high, hampering their practical use.
	Simultaneously, it suffices to approximate either alignments or their corresponding conformance value(s) for many applications.
	Recent work has shown that using subsets of the process model behavior leads to accurate conformance approximations.
	The accuracy of such an approximation heavily depends on the selected subset of model behavior.
	Thus, in this paper, we show that we can derive a priori error bounds for conformance checking approximation based on arbitrary activity sequences, independently of the given process model.
	Such error bounds subsequently let us select the most relevant subset of process model behavior for the alignment approximation.
	Experiments confirm that conformance approximation accuracy improves when using the proposed error bound approximation to guide the selection of relevant subsets of process model behavior.
\end{abstract}
	\section{Introduction}
	\label{sec:intro}

The execution of processes in companies leaves digital event data footprints in the databases of the information systems employed, known as \emph{event logs}.
\emph{Process mining}~\cite{aalst_2016_pm} aims to develop techniques that enhance the overall knowledge of the process by analyzing such event logs, e.g., by automatically discovering process models based on the event log.
\emph{Conformance checking}~\cite{carmona_2018_conformance}, i.e., one of the main sub-fields of process mining, aims at assessing to what degree a given process model and the recorded event data conform to one another. 
In this context, alignments~\cite{adriansyah_2012_alignment}, an established class of conformance checking artifacts, are of particular interest, as they provide an exact quantification of deviations between the recorded process execution and its intended behavior, as modeled by the process model.

The increasing prevalence of information systems in different domains leads to a drastic increase in the amount of recorded event data~\cite{Massi_2018_Big}.
Such \emph{high-volume event data}, combined with complex process models, yield infeasible alignment computation times, hampering their practical use.
Yet, in many applications, exact alignment values are not required, i.e., it suffices to obtain an approximated value to draw meaningful conclusions.
For example, in genetic process discovery~\cite{buijs_2012_role}, various \emph{generations of candidate process models} are evaluated w.r.t. an event log.
Due to the complexity of alignment computation, computing exact alignment results for each candidate process model is impractical.
However, in each generation, rather than obtaining an exact alignment result to judge the process model quality, it is sufficient to know whether a newly generated process model improves its alignment results with respect to that of previous generations.
Therefore, fast alignment approximation techniques, that provide guarantees on the approximation error are of particular interest.

\begin{figure}[tb]
    \centering
    \includegraphics[width=0.75\textwidth]{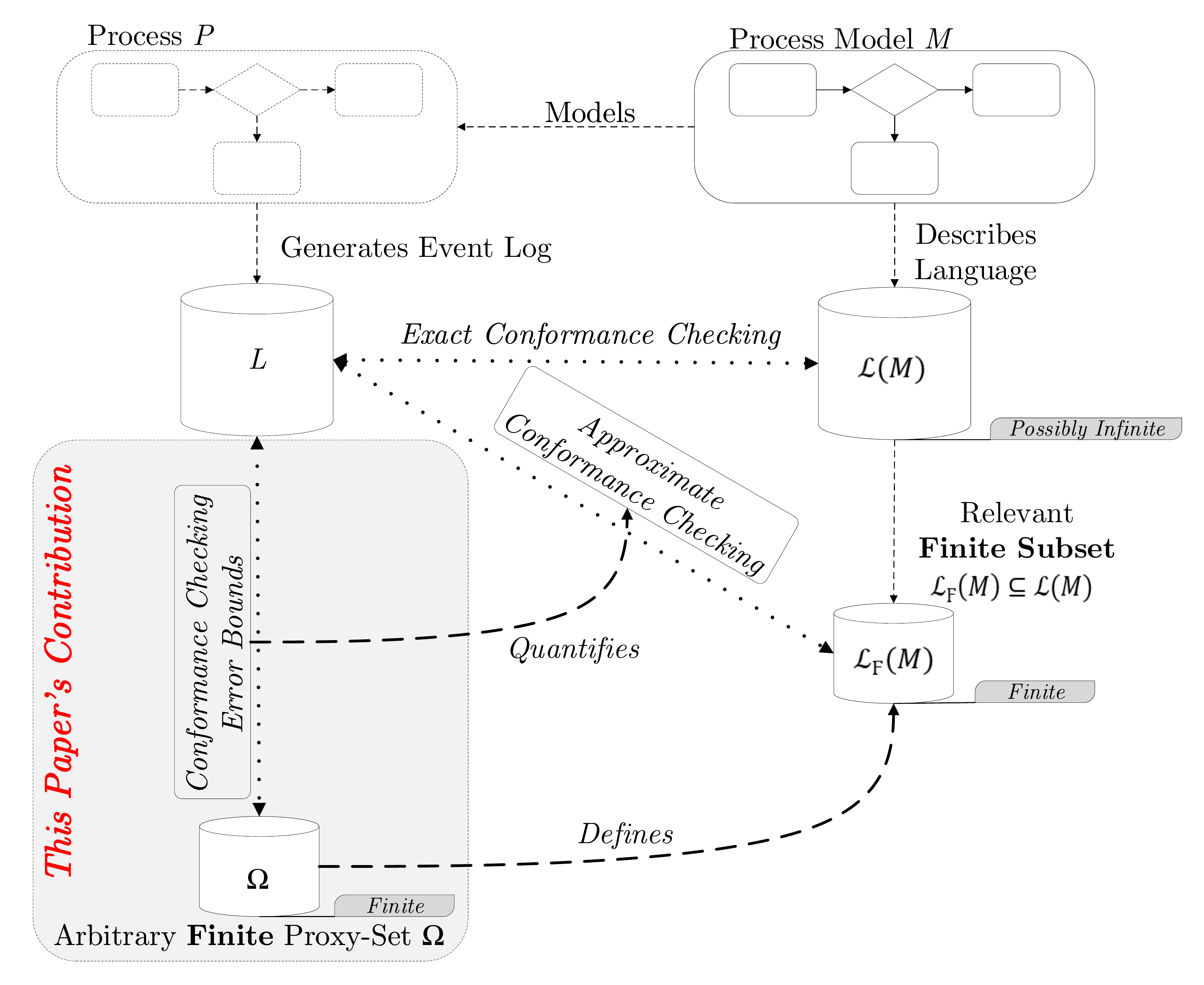}
    \caption{Overview of the proposed approach and its relation to existing work. A process model $\model$ models a process $\process$ that generates an event log $\eventLog$. Existing approaches either compute \emph{exact} or \emph{approximate conformance checking results} based on the language of the model $\lang(\model)$ (or a finite subset thereof). 
    We propose a method that quantifies error bounds for conformance checking approximation, based on an arbitrary proxy-set $\proxySet$.}
    \label{fig:high_level_overview}
\end{figure}

Recently, various approaches for alignment approximation have been proposed~\cite{sani_2020_confsubset,Bas_2017_online}.
In our previous work~\cite{sani_2020_confsubset}, we exploit \emph{subsets of the process model's behavior} for approximation, i.e., by using the subset of process behavior as a representative for the complete process model behavior. 
In this way, we are able to provide bounds for the approximated alignment value.
This branch of approximation techniques shows promising results; however, the approximation result's quality (i.e., the difference between actual and approximated value) heavily depends on the selected subset of process model behavior.
Therefore, quantifying an approximation's quality based on a specific subset of model behavior prior to performing the actual approximation allows us to identify the most appropriate subset to use for said approximation.
In this paper, we present a novel approach to quantify an alignment approximations quality before performing the actual approximation.

Consider \autoref{fig:high_level_overview} (\autopageref{fig:high_level_overview}), in which we present a schematic overview of the approach.
A process model $\model$ models a process $\process$ that generates a digital event log $\eventLog$. 
Existing approaches compute \emph{exact} or \emph{approximate conformance checking results} by considering the language of the model $\lang(\model)$ (possibly infinite), or, a relevant finite subset thereof. 
The proposed method allows us to a-priori compute error bounds for alignment approximation, using an arbitrary proxy-set $\proxySet$, i.e., a set of sequence of activities.
This proxy-set $\proxySet$ can be used to derive the relevant subset of process model behavior $\lang_{F}(\model)$ and may consist of behavior that is not part of the model nor the event log.

We evaluated our error bound estimation technique using various real event logs.
Our experiments confirm a strong correlation between the predicted maximum error bounds and the eventual approximation error.
As such, our experiments confirm that the conformance approximation accuracy improves when using the proposed error bound approximation to guide the selection of relevant subsets of process model behavior.
Furthermore, our experiments show that the error bounds' computation time is negligible w.r.t. computing conventional exact alignments.

The remainder of this paper is structured as follows. 
In \autoref{sec:related_work}, we discuss related work. 
In \autoref{sec:preliminaries}, we present preliminaries and basic notation.
We explain the main methodology of our approach in \autoref{sec:method} and subsequently evaluate it in \autoref{sec:eval}.
Finally, \autoref{sec:conclusion} concludes this work.

\section{Related Work}
\label{sec:related_work}
Several process mining techniques exist, ranging from process discovery to process performance prediction. 
Here, we cover related work in the field of conformance checking and corresponding approximation techniques.
We refer to~\cite{aalst_2016_pm} for an extensive overview and introduction of process mining.

Conformance checking techniques have been well studied.
In \cite{elhagaly_2019_evolution}, the authors review the various conformance checking techniques in the process mining domain.
Similarly, in~\cite{carmona_2018_conformance}, different methods for conformance checking and its applications are covered.
Alignments were introduced in~\cite{van_2012_replaying} and have rapidly developed into the standard conformance checking technique. 
In~\cite{aalst_2013_decomposing,munoz_2014_single}, decomposition techniques are proposed for improving the performance of the alignment computation.
Applying decomposition techniques improves computation time. However, these techniques are able to improve the performance of alignment computation, if there are numerous unique activities in the process~\cite{verbeek_2017_divide}.
Recently, general approximation schemes for alignments, i.e., computation of near-optimal alignments, have been proposed~\cite{taymouri_2016_recursive}.
Finally, the authors in \cite{Bas_2017_online} propose to incrementally compute prefix-alignments, i.e., enabling real-time conformance checking for event data streams.

A limited amount of work considers the use of sampling in process mining. 
%\cite{carmona_2010_parikh} proposed a sampling approach based on Parikh vectors of traces to detect the behavior in the event log.
In~\cite {bauer_2018_much}, the authors recommend a trace-based statistical sampling method to decrease the required discovery time and memory footprint. 
% that assumes the process instances have different behavior if they have different sets of directly follows relations. 
%In~\cite{berti_2017_sampling}, a trace-based sampling method, specifically for the Heuristic miner\cite{weijters_2011_fhm}, is proposed.
%In these sampling methods, we have no control over the size of the final sampled event data, and, depending on the defined behavioral abstraction, the methods may select almost all the process instances.
%Moreover, all these mentioned sampling methods use random trace-based sampling with a replacement that may lead to pick and analyze a unique process instance several times. 
%Lastly, all these sampling methods lead to non-deterministic results. 
Moreover, in \cite{sani_2020_sampling}, we analyzed random and biased sampling methods with which we are able to adjust the size of the sampled data for process discovery.
%Results show that using biased sampling leads to having process models with higher qualities.

Some research has focused on alignment approximation.
In~\cite{nolle_2020_deepalign}, deep learning is used to approximate alignment statistics. 
In~\cite{bauere_2019_stimating}, the authors propose to incrementally sample the event log and applying conformance checking on the sampled data. 
The proposed method increases the sample size until the approximated value is accurate enough. 
The authors of \cite{padro_2019_approximate} propose a conformance approximation method that applies relaxation labeling methods on a partial order representation of a process model, which needs to preprocess the process model each time. 
Unlike these approaches that do not provide bounds for the approximation,  some methods have proposed to generate a subset of model behaviors using instance selection \cite{sani_2020_confsubset} and simulation~\cite{sani_2020_simulation}. 
\cite{sani_2020_confsubset} has proposed to compute alignments of some instances in the event log and use the corresponding model behavior for approximating the alignment of other instances. 

In this paper we prove that by having a subset of model behaviors, we can estimate the approximation error for any process model, thus helping users to adjust the approximation setting. 
Furthermore, we propose some instance selection methods to decrease approximation error.

\section{Preliminaries}
\label{sec:preliminaries}
This section briefly introduces basic conformance checking terminology and the notation used in this paper.
We assume that the reader has a basic knowledge of sets, bags (multisets), Cartesian products functions, and sequences.

We let $\univMSet(X)$ denote the set of all possible bags over $X$.
Given $b{\in}\univMSet(X)$, $\overline{b}{=}\{x{\mid}b(x){>}0\}$.
$X^*$ denotes the set of all sequences over $X$.
Let $X'{\subseteq} X$ and let $\sequence{\in}X^*$, $\sequence_{\downarrow_{X'}}$ returns the projected sequence of $\sequence$ on set $X'$, e.g., $\langle a,b,c,b,d\rangle_{\downarrow_{\{b,d\}}}{=}\langle b,b,d\rangle$.
Let $X_1,X_2,{\dots},X_n$ be $n$ arbitrary sets and let $X_1{\times}X_2{\cdots}{\times}X_n$ denote the corresponding Cartesian product.
Let $\sequence{\in}(X_1{\times}X_2{\cdots}X_n)^*$ be a sequence of tuples, $\pi_i(\sequence)$ returns the sequence of elements in $\sequence$ at position $1{\leq}i{\leq}n$, e.g., $\pi_2(\langle (x^1_1,x_2^1,{\dots},x_n^1),(x^2_1,x_2^2,{\dots},x_n^2),{\dots},(x^{|\sequence|}_1,x_2^{|\sequence|},{\dots},x_n^{|\sequence|})\rangle{=}\langle x^1_2,x^2_2,{\dots}x^{|\sequence|}_2\rangle$.

Given $\sequence,\sequence'{\in}X^*$, $\editDistance(\sequence,\sequence'){\in}\naturals_{\geq0}$ represents the \emph{Longest Common Subsequence (LCS) edit distance} (only using \emph{insertions} and \emph{deletions}) between $\sequence$ and $\sequence'$, i.e., the minimum number of edits required to transform $\sequence$ into $\sequence'$.
For example, $\editDistance(\langle w,x,y\rangle,\allowbreak\langle x,y,z\rangle){=}2$.
Note that $\editDistance(\sequence,\sequence'){=}\editDistance(\sequence',\sequence)$ ($\editDistance$ is symmetrical) and $\editDistance(\sequence,\sequence''){\leq}\editDistance(\sequence,\sequence'){+}\editDistance(\sequence',\sequence'')$ (triangle inequality applies to $\editDistance$).
Given a sequence $\sequence{\in}X^*$ and a set of sequences $S{\subseteq}X^*$, we define $\editDistanceSet(\sequence,S){=}\min\limits_{\sequence'{\in}S}\editDistance(\sequence,\sequence')$. % returns the minimum edit distance value of $\sequence$ w.r.t. all sequences in $S$.

%\subsubsection{Event Log}
\begin{table}[tb]
\caption{Simple example of an event log. Rows capture \emph{events} recorded in the context of the execution of the process. 
%An event describes at what point in time an activity was performed. Other data attributes may be available as well. 
}
\scriptsize
\centering
\resizebox{0.675\textwidth}{!}{%
\begin{tabular}{ |c c c c c c| } 
 \hline
    Case-id & Event-id & Activity name & Starting time & Finishing time &...\\ 
 \hline
    $\vdots$ &$\vdots$ &$\vdots$ &$\vdots$ &$\vdots$ &$\dots$\\ 
    7 & 35 & Register(a) & 2020-01-02 12:23 & 2020-01-02 12:25 &$\dots$\\ 
    7 & 36 & Analyze Defect(b) & 2020-01-02 12:30 & 2020-01-02 12:40 &$\dots$\\ 
    7 & 37 & Inform User(g) & 2020-01-02 12:45 & 2020-01-02 12:47 &$\dots$\\ 
    7 & 38 & Repair(Simple)(c) & 2020-01-02 12:45 & 2020-01-02 13:00 &$\dots$\\ 
    8 & 39 & Register(a) & 2020-01-02 12:23 & 2020-01-02 13:15 &$\dots$\\ 
    7 & 40 & Test Repair(e) & 2020-01-02 13:05 & 2020-01-02 13:20 &$\dots$\\ 
    7 & 41 & Archive Repair(h) & 2020-01-02 13:21 & 2020-01-02 13:22 &$\dots$\\ 
    8 & 42 & Analyze Defect(b) & 2020-01-02 12:30 & 2020-01-02 13:30 &$\dots$\\ 
    %8 & 43 & Inform User(g) & 2020-01-02 12:45 & 2020-01-02 13:47 &$\dots$\\ 
    $\vdots$ &$\vdots$ &$\vdots$ &$\vdots$ &$\vdots$& $\ddots$ \\ 
 \hline
\end{tabular}}
\label{tab:EventLog}
\end{table}
Event logs, i.e., collections of events representing the execution of several instances of a process, are the starting point of process mining algorithms.
An example event log is shown in \autoref{tab:EventLog}.
Events record when an activity was performed (according to their \textit{Starting} and \textit{Finishing time}) for an instance of the process (represented by the \emph{Case-id} column). 
For some applications, e.g., alignment computation, only the \emph{control-flow information}, i.e., sequences of activities executed in the context of a process instance, is required.
Hence, we adopt the aforementioned mathematical model of an event log (in practice however, event data typically records much more features related to the executed activities, e.g.,  resource information and costs of activities).
\begin{definition}[Event Log]
Let $\univAct$ denote the \emph{universe of activities}.
A \emph{trace} $\sequence$ is a sequence of activities ($\sequence{\in}\univAct^*$).
An \emph{event log} $\eventLog{\in}\univMSet(\univAct^*)$ is a bag of traces.
\end{definition}

\begin{figure}[tb]
	\centering
	\includegraphics[width=0.95\textwidth]{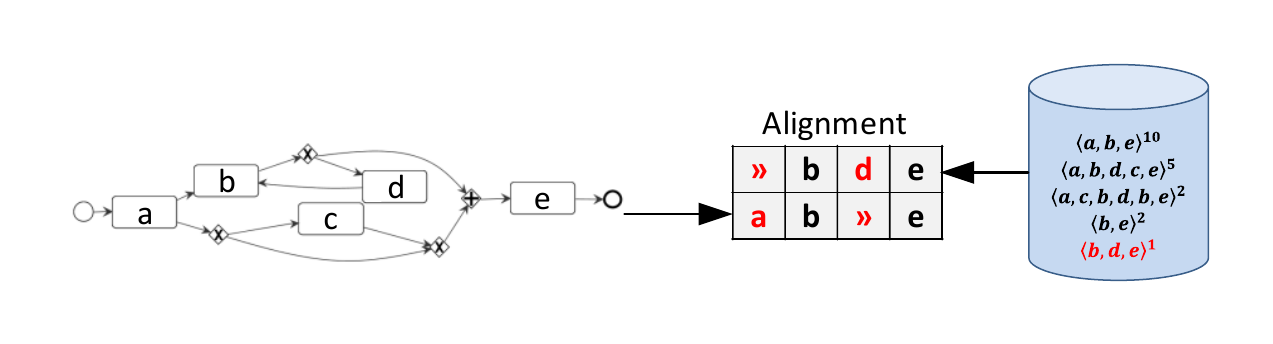}
		\caption{An example of a process model $\model_1$ (in BPMN notation) and a simple event log $\eventLog_1$ (represented in multiset-view). 
		The optimal alignment of the last trace of the event log and the process model is shown in the middle of the figure. 
		}
		\label{fig:bpmn}
\end{figure}

Process models are used to describe the (expected) behavior of a process.
Process models come in various forms, i.e., ranging from simple conceptual drawings to mathematical concepts with associated execution semantics, e.g., Petri nets~\cite{petri_2008_petri} or BPMN diagrams~\cite{BPMN}.
For example, in \autoref{fig:bpmn}, we show process model $\model_1$ in BPMN notation.
The model describes that the first activity in the process should be \textit{a}, followed by activities \textit{b} and \textit{c} are in parallel. 
It is possible to skip activity \textit{c}. 
After the execution of activity \textit{b}, if we execute activity \textit{d}, we should again execute \textit{b}. 
Activity \textit{e} finalizes the process.
In this paper's context, we do not assume a specific modeling notation; rather, we assume process models to describe a collection of sequences of activities.

\begin{definition}[Process Model]
Let $\univAct$ denote the universe of activities.
A process model $\model$ describes the intended behavior of a process.
We refer to the behavior described by model $\model$ as its \emph{language} $\emptyset{\subset}\lang(\model){\subseteq}\univAct^*$, i.e., a non-empty collection of activity sequences.
\end{definition}
For the given process model $\model_1$ in \autoref{fig:bpmn}, we have $\lang(\model_1){=}\{\langle a,b,e\rangle, \langle a,b,c,e\rangle, \langle a,c,b,e\rangle, \langle a,b,d,b,e\rangle, {\dots} \}$.
Note that, due to the existence of loops, the language of a process model may be infinite.

To quantify whether an event log conforms to a process model, we use alignments. 
An alignment between a trace and a model describes which events in the trace can be \enquote{aligned with activities described by the process model}.
Furthermore, alignments indicate whether an event cannot be explained by the model or whether an activity as described by the model was not observed.
In \autoref{fig:bpmn}, an alignment of trace $\langle b,d,e\rangle$, and the given process model is provided.
Observe that the trace does not contain activity $a$, which should always be present according to the model.
In the alignment, this is visualized by the first column $\frac{\gg}{a}$.
Similarly, after the observed $d$-activity, no second $b$-activity was observed.
As such, in this alignment, the occurrence of $d$ is rendered obsolete, i.e., visualized as $\frac{d}{\gg}$.
We formally define alignments as follows.

\begin{definition}[Alignment]
Let $\univAct$ denote the universe of activities, let $\model$ be a process model with corresponding language $\emptyset{\subset}\lang(\model){\subseteq}\univAct^*$ and let $\sequence{\in}\univAct^*$ be a trace.
An alignment $\alignment$ of $\sequence$ and $\model$, is a sequence, characterized as $\alignment{\in}((\univAct{\cup}\{\gg\}){\times}(\univAct{\cup}\{\gg\}))^*$, s.t., $\pi_1(\alignment)_{\downarrow_{\univAct}}{=}\sequence$ and $\pi_2(\alignment)_{\downarrow_{\univAct}}{\in}\lang(\model)$.
The set of all possible alignments of trace $\sequence$ and model language $\lang(\model)$ is denoted as $\univAlignments(\sequence,\lang(\model))$.
\end{definition}
In the context of this paper, given $\alignment{\in}\univAlignments(\sequence,\lang(\model))$, we write  $\modelTrace(\alignment){=}\pi_2(\alignment)_{\downarrow_{\univAct}}$ to refer to \emph{the \enquote{model trace} corresponding to $\alignment$}.
Let $\costOfMove{\colon}(\univAct{\cup}\{\gg\}){\times}(\univAct{\cup}\{\gg\}){\to}\reals$, then, given $\sequence{\in}\univAct^*$, $\model{\subseteq}\univAct^*$ and $\alignment{\in}\univAlignments(\sequence,\lang(\model))$, we let $\costOfAlignmentFunc_{\costOfMove}(\alignment){=}\sum\limits_{1{\leq}i{\leq}|\alignment|}\costOfMove(\alignment(i))$ denote the cost of alignment $\alignment$.
We let $\univAlignments_{\costOfMove}^{\star}(\sequence,\lang(\model)){=}\argmin\limits_{\alignment{\in}\univAlignments(\sequence,\lang(\model))}\costOfAlignmentFunc_{\costOfMove}(\alignment)$ be the set of \emph{optimal/minimal alignments}.
We let $\optimalAlignmentCost_{\costOfMove}(\sequence,\lang(\model)){=}\min\limits_{\alignment{\in}\univAlignments(\sequence,\lang(\model))}\costOfAlignmentFunc_c(\alignment)$ be the optimal alignment cost for trace $\sequence$ and model $\model$ (hence: $\forall{\alignment{\in}\univAlignments_{\costOfMove}^{\star}}(\sequence,\lang(\model))\left(\costOfAlignmentFunc_{\costOfMove}(\alignment){=}\optimalAlignmentCost_{\costOfMove}(\sequence,\lang(\model)) \right)$).
In the remainder, we assume that $c$ represents the \emph{standard cost function}, i.e., $c(a,\gg){=}c(\gg,a){=}1,\forall a{\in}\univAct$, $c(a,a){=}0,\forall a{\in}\univAct$ and $c(a,a'){=}\infty, \forall a{\neq}a'{\in}\univAct$, and we omit it as a subscript.
%Since $\univAlignments$, $\univAlignments^{\star}$ and $\optimalAlignmentCost$ take a sequence and collection of sequences as an input, given an arbitrary sequence $\sequence{\in}\univAct^*$ and non-empty set of sequences $S{\subseteq}\univAct^*$, $\univAlignments(\sequence,S)$, $\univAlignments^{\star}(\sequence,S)$ and $\optimalAlignmentCost(\sequence,S)$ are readily defined.

%%%%%%%%%%%%%%%%%%%%%% Section 4$$$$$$$$$$$$$$$$$$
\section{Alignment Error Bound Estimation}
\label{sec:method}
In this section, we show how to estimate maximal alignment approximation error bounds.
We first show that \emph{edit distance} can be used to compute conventional optimal alignments.
Subsequently, we use this result to show that we are able to quantify the maximum alignment approximation error for an arbitrarily given activity sequence.
Finally, we show that we can guarantee tighter approximation bounds by exploiting a collection of arbitrary activity sequences.
%Finally, we provide some instance selection methods to generate an event log's proxy-set. 

%\subsection{ Optimal Alignments }
\subsection{Computing the Maximal Alignment Approximation Error}
In this section, we show that, for given traces $\sequence,\sequence'{\in}\univAct^*$ and process model $\model$, $\editDistanceSet(\sequence,\sequence')$ quantifies a range for the optimal alignment value $\optimalAlignmentCost(\sequence,\lang(\model))$, when using $\optimalAlignmentCost(\sequence',\lang(\model))$ as an estimator for $\optimalAlignmentCost(\sequence,\lang(\model))$. 
We first show that for the standard cost function, we are able to use the LCS edit distance function to compute the cost of the optimal alignment. 
\begin{lemma}[Edit Distance Quantifies Alignment Costs]
\label{lemma:edit_distance_single_alignment_costs}
Let $\univAct$ denote the universe of activities, let $\sequence{\in}\univAct^*$ be a trace, let $\model$ be a process model and let $\alignment{\in}\univAlignments^{\star}(\sequence,\lang(\model))$ be an optimal alignment of $\sequence$ and $\model$.
Using the standard cost function, 
$\costOfAlignmentFunc(\alignment){=}\editDistance(\sequence,\modelTrace(\alignment))$
\begin{proof}
Observe that $\alignment$ only contains elements of the form $(a,a)$, $(a,\gg)$ and $(\gg,a)$.
Let $R$ denote the set of elements of the form $(a,\gg)$ and let $I$ denote the set of elements of the form $(\gg,a)$.
Transforming $\sequence$ into $\modelTrace(\alignment)$ is achieved by \emph{removing} activities in $\sequence$ represented by the elements in $R$ and inserting activities in $\sequence$ represented by the elements in $I$.
Hence, $\costOfAlignmentFunc(\alignment){=}R{+}I$.
Similarly, $\editDistance(\sequence,\modelTrace(\alignment))$ represents the {minimum number of insertions and removals} to transform $\sequence$ into $\modelTrace(\alignment)$.
Thus, if $\costOfAlignmentFunc(\alignment){<}\editDistance(\sequence,\modelTrace(\alignment))$, then $\editDistance(\sequence,\modelTrace(\alignment))$ does not represent the minimal number of edits.
Similarly, if $\costOfAlignmentFunc(\alignment){>}\editDistance(\sequence,\modelTrace(\alignment))$ then $\alignment$ is not optimal. $\hfill \square$
\end{proof}
\end{lemma}

\begin{corollary}[$\editDistanceSet(\sequence,\lang(\model))$ equals $\optimalAlignmentCost(\sequence,\lang(\model))$]
\label{cor:edit_ditance_global_alignment_costs}
Let $\univAct$ denote the universe of activities, let $\sequence{\in}\univAct^*$ be a trace, let $\model$ be a process model with corresponding language $\emptyset{\subset}\lang(\model){\subseteq}\univAct^*$.
Using the standard cost function, $\optimalAlignmentCost(\sequence,\lang(\model)){=}\editDistanceSet(\sequence,\lang(\model))$.
\begin{proof}
Let $\alignment{\in}\univAlignments^{\star}(\sequence,\lang(\model))$.

$\optimalAlignmentCost(\sequence,\lang(\model)){=}\costOfAlignmentFunc(\alignment){=}\editDistance(\sequence,\modelTrace(\alignment)){=}\editDistanceSet(\sequence,\lang(\model))$. $\hfill \square$
\end{proof}
\end{corollary}

In the following, we show that it is possible to exploit an arbitrary activity sequence to derive a range on another activity sequence's alignment value.
\begin{theorem}[Edit Distance Provides Approximation Bounds]
\label{th:bounds}
Let $\sequence,\sequence'{\in}\univAct^*$ be two traces and let $\model$ be a process model with corresponding language $\emptyset{\subset}\lang(\model){\subseteq}\univAct^*$.
The optimal alignment value $\optimalAlignmentCost(\sequence,\lang(\model))$, is within $\editDistance(\sequence,\sequence')$ of $\optimalAlignmentCost(\sequence',\lang(\model))$, i.e., 
$\optimalAlignmentCost(\sequence',\lang(\model)){-}\editDistance(\sequence,\sequence'){\leq}\optimalAlignmentCost(\sequence,\lang(\model)){\leq}\optimalAlignmentCost(\sequence',\lang(\model)){+}\editDistance(\sequence,\sequence')$.
\begin{proof}
Let $\alignment{\in}\univAlignments^{\star}(\sequence,\lang(\model))$ and let $\alignment'{\in}\univAlignments^{\star}(\sequence',\lang(\model))$.
Triangle inequality of LCS edit distance yields
$\editDistance(\sequence,\modelTrace(\alignment')){\leq}\editDistance(\sequence,\sequence'){+}\editDistance(\sequence',\modelTrace(\alignment'))$, which we can rewrite ( Lemma~\ref{lemma:edit_distance_single_alignment_costs}) to $\editDistance(\sequence,\modelTrace(\alignment')){\leq}\editDistance(\sequence,\sequence'){+}\optimalAlignmentCost(\sequence',\lang(\model))$.
Since $\optimalAlignmentCost(\sequence,\lang(\model)){\leq}\allowbreak\editDistance(\sequence,\modelTrace(\alignment'))$, we have: $\optimalAlignmentCost(\sequence,\lang(\model)){\leq}\optimalAlignmentCost(\sequence',\lang(\model)){+}\editDistance(\sequence,\sequence')$.

Similarly, $\editDistance(\sequence',\modelTrace(\alignment)){\leq}\editDistance(\sequence,\sequence'){+}\editDistance(\sequence,\modelTrace(\alignment))$.
Again, we deduce $\editDistance(\sequence',\modelTrace(\alignment)){\leq}\allowbreak\editDistance(\sequence,\sequence'){+}\optimalAlignmentCost(\sequence,\lang(\model))$.
Since $\optimalAlignmentCost(\sequence',\lang(\model)){\leq}\editDistance(\sequence',\modelTrace(\alignment))$, we deduce $\optimalAlignmentCost(\sequence',\lang(\model)){-}\editDistance(\sequence,\sequence'){\leq}\allowbreak\optimalAlignmentCost(\sequence,\lang(\model))$. 
Hence, we obtain $\optimalAlignmentCost(\sequence',\lang(\model)){-}\editDistance(\sequence,\sequence'){\leq}\allowbreak\optimalAlignmentCost(\sequence,\lang(\model)){\leq}\optimalAlignmentCost(\sequence',\lang(\model)){+}\editDistance(\sequence,\sequence')$. $\hfill \square$
\end{proof}
\end{theorem}
For example, reconsider process model $\model_1$ and event log $\eventLog_1$ in \autoref{fig:bpmn}.
Observe that $\optimalAlignmentCost(\langle a,c,c,b,d,e\rangle, \lang(\model_1)){=}2$ and $\editDistance(\langle a,c,c,b,d,e\rangle,\langle a,c,b,d,e\rangle){=}1$.
Hence, we deduce $1{\leq}\optimalAlignmentCost(\langle a,c,b,d,e\rangle, \lang(\model_1)){\leq}3$.
If $\optimalAlignmentCost(\langle a,c,c,b,d,e\rangle, \lang(\model_1))$ is unknown, $\editDistance(\langle a,c,c,b,d,e\rangle,\langle a,c,b,d,e\rangle){=}1$ implies that using it as an estimator for $\optimalAlignmentCost(\langle a,c,b,d,e\rangle, \lang(\model_1))$ \emph{yields a maximal absolute approximation error of $1$}.

\subsection{Generating Proxy-Sets}
\label{subsec:instanceselection}
\begin{comment}
\begin{definition}[Alignment Approximation Error Bound]
\label{def:approximateEB}
Let $\univAct$ denote the universe of event labels, let $\model$ be a process model, and let $\proxySet{\subseteq}\univAct^*$ be an arbitrary set of activity sequences.
We define $\approxError_{\proxySet} {\colon} \univAct^* {\to} \naturals$ as the $\proxySet$-driven alignment approximation error bound where $\approxError_{\proxySet}(\sequence){=}\editDistanceSet(\sequence,\proxySet)$.
\end{definition}
\end{comment}
The result of \autoref{th:bounds} implies that, given a process model $\model$ and traces $\sequence,\sequence'{\in}\univAct^*$, when using $\optimalAlignmentCost(\sequence',\lang(\model))$ as an estimator for $\optimalAlignmentCost(\sequence,\lang(\model))$, we obtain an approximation error $\approxError{\leq}\editDistance(\sequence,\sequence')$.
Interestingly, the bound on $\approxError$ is determined independently of the model.
Furthermore, $\sequence'$ is allowed to be an arbitrary sequence, i.e., it is perfectly fine if $\sequence'{\notin}\lang(\model)$, and, given some $\eventLog{\in}\univMSet(\univAct^*)$ s.t. $\sequence{\in}\overline{\eventLog}$, $\sequence'{\notin}\overline{\eventLog}$.
Hence, given an arbitrary set of sequences $\proxySet{\subseteq}\univAct^*$, $\argmin\limits_{\sequence'{\in}\proxySet}\editDistance(\sequence,\sequence')$ represents the members of $\proxySet$ that minimize the expected maximum error when using $\optimalAlignmentCost(\sequence',\lang(\model))$ as an estimator (i.e., for $\sequence'{\in}\argmin\limits_{\sequence'{\in}\proxySet}\editDistance(\sequence,\sequence')$).
In the remainder, we refer to such a set of sequences $\proxySet{\subseteq}\univAct^*$ as a \emph{proxy-set}, i.e., we intend to \enquote{align by proxy through $\proxySet$}.

Observe that, for an event log $\eventLog{\in}\univMSet(\univAct^*)$ and proxy-set $\proxySet{\subseteq}\univAct^*$, $\forall{\sequence{\in}\overline{\eventLog}}\left(\min\limits_{\sequence'{\in}\proxySet}\editDistance(\sequence,\sequence'){=}0\right){\Leftrightarrow}\proxySet{\supseteq}\overline{\eventLog}$, i.e., if every member of the log has an edit distance of $0$ w.r.t. the proxy-set, then every member of the event log is a member of the proxy-set (and vice versa).
Clearly, in such a case, using proxy-set $\proxySet$ yields optimal alignments, yet, at the same (or even worse) time and memory complexity as computing conventional optimal alignments.

%Hence, we are primarily interested in a proxy-set $\proxySet$ that is significantly smaller than the original event log.
In the remainder, given an event log  $\eventLog{\in}\univMSet(\univAct^*)$ and proxy-set $\proxySet{\subseteq}\univAct^*$, we let $\epsilon_{\proxySet}(\eventLog){=}\sum\limits_{\sequence{\in}\overline{\eventLog}}\eventLog(\sequence){\cdot}\min\limits_{\sequence'{\in}\proxySet}\editDistance(\sequence,\sequence')$.
Given two proxy-sets $\proxySet,\proxySet'{\subseteq}\univAct^*$, $\proxySet$  \emph{dominates} $\proxySet'$ for event log $\eventLog$ if and only if $\approxError_{\proxySet}(\eventLog){\leq}\approxError_{\proxySet'}(\eventLog)$ and $|\proxySet|{<}|\proxySet|'$.
In such a case, we refer to $\proxySet'$ as a \emph{redundant} proxy-set.
A proxy-set $\proxySet$ is \emph{$k$-optimal} for event log $\eventLog$ if and only if $\forall{\proxySet'{\in}\univAct^*}\left(|\proxySet'|{=}k{\implies}\approxError_{\proxySet}(\eventLog){\leq}\approxError_{\proxySet'}(\eventLog)\right)$.
A \emph{$k$-optimal} proxy-set $\proxySet$ is \emph{$k$-primal} if $|\proxySet|{=}k$.
For example, $\proxySet{=}\overline{\eventLog}$ is $|\overline{L}|$-primal, $1$-optimal, $2$-optimal, ${\dots}$, $|\overline{L}|$-optimal.
Furthermore, it is easy to see that any ($k$-primal) proxy-set $\proxySet$ with $|\proxySet|{>}\eventLog$ is dominated by $\eventLog$ and hence redundant.
More interestingly, primal proxy-sets that are smaller than the event log are never redundant.
\begin{theorem}[Primal Proxy-Sets are Non-Redundant]
\label{th:primal_non_redundant}
Let $\eventLog{\in}\univMSet(\univAct^*)$ be an event log and let $\proxySet{\subseteq}\univAct^*$ be a proxy-set such that $|\proxySet|{<}|\overline{\eventLog}|$ and $\proxySet$ is \emph{k-primal}.
$\proxySet$ is \emph{non-redundant}.
\begin{proof}
Assume that $\proxySet$ is redundant.
Hence, $\exists{\proxySet'{\subseteq}\univAct^*}\left(|\proxySet'|{<}|\proxySet|{\wedge}\approxError_{\proxySet'}(\eventLog){\leq}\approxError_{\proxySet}(\eventLog) \right)$.
However, observe that, we are able to create $\proxySet''{=}\proxySet'{\cup}\eventLog''$ with $|\eventLog''|{=}|\proxySet|{-}|\proxySet'|$ and $\sequence{\in}\eventLog''{\implies}\sequence{\in}\overline{\eventLog}{\wedge}\sequence{\notin}\proxySet'$ (note that $|\proxySet|{=}|\proxySet''|$).
Observe that $\approxError_{\proxySet''}(\eventLog){<}\approxError_{\proxySet'}(\eventLog)$ and as a consequence $\approxError_{\proxySet''}(\eventLog){<}\approxError_{\proxySet}(\eventLog)$, contradicting the fact that $\proxySet$ is $k$-primal. $\hfill \square$
\end{proof}
\end{theorem}

Observe that \autoref{th:primal_non_redundant} implies that for any event log $\eventLog{\in}\univMSet(\univAct^*)$ and $k{\in}1,2,{\dots}|\eventLog|$, there exists a $k$-primal proxy-set $\proxySet$.
A $k$-primal proxy-set minimizes the maximal possible error bound, and hence, can be regarded as the \emph{optimal} proxy-set to use of size $k$.
However, providing such proxy-sets is usually NP-Hard.
In the upcoming paragraphs, we briefly introduce various proxy-set generation methods and their relation to primal proxy-sets.

\subsubsection{Sampling}
Proxy-sets can be generated using sampling methods: either sampling members for the input event log, the given process model, or a mixture thereof.
In previous work, we investigated sampling of model behavior using uniform distributions~\cite{sani_2020_confsubset} and event-log-guided process model simulation~\cite{sani_2020_simulation}.

Strictly sampling the behavior from the process model, i.e., $\proxySet{\subseteq}\lang(\model)$, particularly when using event-log-guided simulation yields (under standard cost function) $\optimalAlignmentCost(\sequence',\lang(\model)){=}0,\ \forall{\sequence'{\in}\proxySet}$. 
On the one hand, it is very unlikely that such a proxy-set is $k$-primal.
On the other hand, the fact that $\optimalAlignmentCost(\sequence',\lang(\model)){=}0,\ \forall{\sequence'{\in}\proxySet}$, can be exploited.
For example, given some trace $\sequence{\in}\overline{\eventLog}$ and some $\sequence'{\in}\argmin\limits_{\sequence'{\in}\proxySet}\editDistance(\sequence,\sequence')$, rather than using $\optimalAlignmentCost(\sequence',\lang(\model))$ (i.e., value $0$) as an estimator for $\optimalAlignmentCost(\sequence,\lang(\model))$, one can use $\frac{\editDistance(\sequence,\sequence')}{2}$.
Hence, the maximal approximation error is reduced by half.

Sampling $\proxySet$ from the event log is likely to result in a proxy-set that is closer to a $k$-primal solution, i.e., in particular when prioritizing sampling of $\sequence{\in}\overline{\eventLog}$ with high $\eventLog(\sequence)$ values.
Hence, using event log-based sampling typically yields smaller values for the maximal obtainable approximation error.
However, since the actual $\optimalAlignmentCost(\sequence',\lang(\model))$ for $\sequence'{\in}\proxySet$ is unknown, we cannot tighten the estimator.

\subsubsection{Centroid-Based Clustering}
For a given target size $k$, the best possible obtainable proxy-set is \emph{$k$-primal}.
As an alternative approach to sampling, \emph{clustering algorithms}~\cite{xu2015comprehensive} are a suitable proxy-set selection mechanism.
A clustering algorithm groups a set of objects into subgroups (clusters) such that the members of a cluster are similar/close, given some similarity/distance metric.
In the case of proxy-set generation, the edit distance serves as a distance metric.
\emph{Centroid-based clustering algorithms}, i.e., algorithms that define clusters using a \emph{central object} (the centroid), are of particular interest.
In centroid-based clustering, the algorithms assign the objects to the centroid objects that are at a minimal distance of the object.
As an example, the \emph{K-Medoids} algorithm~\cite{DBLP:journals/eswa/ParkJ09} uses objects from the object set as centroids and minimzizes the pair-wise dissimilarity of the objects and the centroids.
Clearly, several variations of centroid-based clustering algorithms can be used.
Whereas the clustering algorithms can be applied on an arbitrary set of activity sequences, applying them on the input event log yields proxy-sets that are close to the \emph{$k$-primal} solution.

\subsection{Deriving Exact Alignment Approximation Bounds}
Thus far, given sequence $\sequence{\in}\univAct^*$ and process model $\model$, we have shown that a proxy-set $\proxySet$ and proxy-sequence $\sequence'{\in}\proxySet$ quantify the \emph{maximum approximation error}, when using $\optimalAlignmentCost(\sequence',\lang(\model))$ as an estimator for $\optimalAlignmentCost(\sequence,\lang(\model))$.
In this section, we show that we can exploit proxy-sets to derive \emph{exact approximation bounds}.

When approximating alignments using proxy-set $\proxySet$, we first compute the alignments of the proxy-set traces (i.e., $\sequence'{\in}\proxySet$).
We derive the upper and lower bound of the alignment cost of $\optimalAlignmentCost(\sequence,\lang(\model))$ by simply adding/subtracting $\editDistance(\sequence,\sequence')$ to $\optimalAlignmentCost(\sequence',\lang(\model))$.
Observe that, when using the standard cost function, the lower bound of any alignment is bounded, i.e., it cannot be lower than $0$.
%For the upper-bound, such a bound does not exist.
Furthermore, in certain cases, it is possible to derive a tighter lower-bound.
Let $\univAct_{\model}{=}\{a{\in}\univAct{\mid}\exists{\sequence{\in}\lang(\model)}(a{\in}\sequence)\}$, then, for any $\sequence{\in}\univAct^*$, it is easy to see that $\optimalAlignmentCost(\sequence,\lang(\model)){\geq}|\sequence_{\downarrow_{\univAct{\setminus}\univAct_{\model}}}|$, i.e., the elements of $\sequence_{\downarrow_{\univAct{\setminus}\univAct_{\model}}}$ are always moves of the form $\frac{a}{\gg}$.
Furthermore, in case $|\sequence|{<}\min\limits_{\sequence'{\in}\lang(\model)}|\sequence'|$, we need at least $|\sequence'|{-}|\sequence|$ (where $\sequence'{\in}\argmin\limits_{\sequence'{\in}\lang(\model)}|\sequence'|$) moves of the form $\frac{\gg}{a}$.
%\footnote{For any $\sequence{\in}\univAct^*$ and $\sequence'{\in}\argmin\limits_{\sequence'{\in}\lang(\model)}|\sequence'|$, we always need at least $|\sequence'|{-}|\sequence|$ moves, however, only when $|\sequence|{\leq}|\sequence'|$ then $|\sequence'|{-}|\sequence|{\geq}0$.}
%Hence, the theoretical lower-bound of any $\sequence{\in}\univAct^*$ is equal to $\max(0,|\sequence|-\min\limits_{\sequence'{\in}\lang(\model)}|\sequence'|){+}|\sequence_{\downarrow_{\univAct{\setminus}\univAct_{\model}}}|$.
Hence, the theoretical lower-bound of any $\sequence{\in}\univAct^*$ is equal to $\max(0,\min\limits_{\sequence'{\in}\lang(\model)}(|\sequence'|){-}|\sequence|){+}|\sequence_{\downarrow_{\univAct{\setminus}\univAct_{\model}}}|$.
We correspondingly define the $\proxySet$-driven lower and upper bound as follows.
\begin{definition}[$\proxySet$-Driven Alignment Bounds]
Let $\univAct$ denote the universe of activities, let $\model$ be a process model with corresponding language $\emptyset{\subset}\lang(\model){\subseteq}\univAct^*$ and let $\proxySet{\subseteq}\univAct^*$ be a proxy-set.
We let $\ubo_{\proxySet,\model}{\colon}\univAct^*{\to}\naturals$ denote the $\proxySet$-driven upper bound and we let $\lobo_{\proxySet,\model}{\colon}\univAct^*{\to}\naturals$ denote the $\proxySet$-driven lower bound where:
\begin{equation}\small
    \label{eq:ubo_proxy}
    \ubo_{\proxySet,\model}(\sequence){=}\min\limits_{\sequence'{\in}\proxySet}(\optimalAlignmentCost(\sequence',\lang(\model)){+}\editDistance(\sequence,\sequence'))
\end{equation}
\begin{equation}\small
    \label{eq:lobo_proxy}
    \lobo_{\proxySet,\model}(\sequence){=}\max(\max(0,\min\limits_{\sequence'{\in}\lang(\model)}(|\sequence'|){-}|\sequence|){+}|\sequence_{\downarrow_{\univAct{\setminus}\univAct_{\model}}}|,\max\limits_{\sequence'{\in}\proxySet}(\optimalAlignmentCost(\sequence',\lang(\model)){-}\editDistance(\sequence,\sequence')))
\end{equation}
\end{definition}
Finally, given $\ubo_{\proxySet,\model}$ and $\lobo_{\proxySet,\model}$, we quantify the alignment approximation value of $\sequence{\in}\univAct^*$, i.e., $\hat{\optimalAlignmentCost}_{\proxySet}(\sequence, \lang(\model))$, as $\hat{\optimalAlignmentCost}_{\proxySet}(\sequence, \lang(\model)){=}\frac{\ubo_{\proxySet,\model}(\sequence){-}\lobo_{\proxySet,\model}(\sequence)}{2}$.
Theoretically, it is possible to give different weights to lower and upper bounds. However, finding the best weight is not the scope of this paper.

\section{Evaluation}
\label{sec:eval}

In this section, we explore the accuracy and the performance of our proposed method. 
First, we briefly describe the implementation, after which we explain the experimental setting.
Finally, we report on the experimental results.

\subsubsection{Implementation}
To apply the proposed conformance approximation method, we implemented the \textit{Conformance Approximation} plug-in in the \texttt{ProM}~\cite{van_2009_prom} framework\footnote{\small \url{svn.win.tue.nl/repos/prom/Packages/LogFiltering}}, including various proxy-set generation methods (both sampling and centroid-based clustering, cf. \autoref{subsec:instanceselection}).
%In addition, to apply the methods on various event logs with different parameters, we ported the developed plug-in to \texttt{RapidProM}, i.e., an extension of \texttt{RapidMiner} and combines scientific work-flows with several process mining algorithms~\cite{Rad}.

\subsection{Experimental Setup}
\label{subsec:ES}
We applied the proposed methods to six real event logs. 
Basic information, e.g., the number of distinct activities, traces, and variants, of the event logs used, is given in \autoref{tab:EventLOgs}.
\begin{table*}[tb]
	\centering
	\caption{Statistics regarding the real event logs that are used in the experiment.}
	\label{tab:EventLOgs}
	\scriptsize
	\begin{tabular}{|l|c|c|c|c|}
		\hline
		\textbf{Event Log} & \textbf{Activities} & \textbf{Traces} & \textbf{Variants} & \textbf{DF\#} \\ \hline
		\textit{$ \text{BPIC-}2012$}~\cite{BPI_2012_challeng}   & 23 & 13087 & 4336 & 138\\ \hline
		%\textit{$ \text{BPIC-}2018\text{-Department}$}~\cite{BPI_2018_challeng} & 6 & 29297 & 349 & 19 \\ \hline
		\textit{$ \text{BPIC-}2018\text{-Inspection}$}~\cite{BPI_2018_challeng}  & 15 & 5485 & 3190 & 67    \\ \hline
		%\textit{$ \text{BPIC-}2018\text{-Reference}$}~\cite{BPI_2018_challeng}  & 6 & 43802 & 515 & 15  \\ \hline
		\textit{$ \text{BPIC-}2019 $}~\cite{BPI_2019_challeng}   & 42 & 251734 & 11973 & 498   \\ \hline
		\textit{$\text{Hospital-Billing} $}~\cite{mannhardt_2017_hospital}  & 18 & 100000 & 1020 & 143  \\ \hline
		\textit{$ \text{Road} $}~\cite{de_2015_road}   & 11 & 150370 & 231 & 70  \\ \hline
		\textit{$ Sepsis $}~\cite{Sepcis_2016_Felix}  & 16 & 1050 & 846 & 115 \\ \hline
	\end{tabular}

\end{table*}
For each event log, we apply conformance checking using different process models.
To obtain these process models, we used the Inductive Miner~\cite{leemans_2014_ind_infreq} process discovery algorithm, with infrequent thresholds equal to $ 0.2 $, $ 0.4 $, and $ 0.6 $.
Typically, these models describe a strict subset of the input event log.
We used four different proxy-set generation methods, all using the event log as a primary driver, i.e., \emph{random sampling}, \emph{frequency-based sampling}, \emph{K-Medoids clustering} and \emph{K-Center clustering}.
In random sampling, we randomly sample variants (without replacement) from the event log to act as a proxy.
In frequency-based sampling, we select traces based on their $\eventLog(\sequence)$-values, in descending order.
In K-Medoids clustering, centroids are determined by minimizing the pair-wise dissimilarity.
In K-Center clustering, the maximum distance between centroids and the objects is minimized.
To determine the size of proxy-sets we use a different percentage of the number variants in the event logs, i.e., $5\%$, $\%10$, $20\%$, $30\%$, $50\%$. 
Moreover, we have repeated each experiment four times as some results are non-deterministic.

Using the described experimental setup, we investigate the relationship between the maximum error,i.e., $\sum\limits_{\sequence\in L}(L(\sequence){\times}\min\limits_{\sequence'{\in}\proxySet}\editDistance(\sequence,\sequence'))$ and the eventual approximation error, as well as the performance of the approach.

\subsection{Results}
In this section, we discuss the results of the experiments.
We first investigate the relationship between the estimated maximum error and the actual approximation error.
Secondly, we investigate the time-performance of the estimation.
Lastly, we investigate the role of the new proposed lower bound. 
\subsubsection{Estimated Maximum Error versus Approximation Error}
Observe that minimizing the expected maximum error does not guarantee a minimal approximation error.
For example, given some model $\model$, $\sequence{\in}\univAct^*$, $\proxySet{=}\{\sequence_1, \sequence_2\}$ and $\proxySet'{=}\{\sequence_1,\sequence_3\}$, assume that $\editDistance(\sequence,\sequence_1){=}2$, $\editDistance(\sequence,\sequence_2){=}3$ and $\editDistance(\sequence,\sequence_3){=}1$.
Clearly, the maximal error based on $\proxySet$ is $2$, and, based on $\proxySet'$, it is $1$.
As such, we intuitively favor $\proxySet'$ over $\proxySet$.
However, if $\optimalAlignmentCost(\sequence_1,\lang(\model)){=}7$, $\optimalAlignmentCost(\sequence_2,\lang(\model)){=}2$ and $\optimalAlignmentCost(\sequence_1,\lang(\model)){=}6$, we obtain $\lobo_{\proxySet,\model}(\sequence){=}\ubo_{\proxySet,\model}(\sequence){=}5$, whereas $\lobo_{\proxySet',\model}(\sequence){=}5$ and $\ubo_{\proxySet',\model}(\sequence){=}7$.
Hence, from $\proxySet$, we derive that $\optimalAlignmentCost(\sequence,\lang(\model)){=}5$ (note $\hat{\optimalAlignmentCost}_{\proxySet}(\sequence,\lang(\model)){=}5$), whereas from $\proxySet'$, we derive $5{\leq}\optimalAlignmentCost(\sequence,\lang(\model)){\leq}7$ (with $\hat{\optimalAlignmentCost}_{\proxySet'}(\sequence,\lang(\model)){=}6$).
Thus, using $\proxySet$ to derive the alignment approximation value actually yields the exact alignment value using $\proxySet'$ yields an error of $1$.

\begin{figure}[tb]
    \centering
    \includegraphics[width=\textwidth]{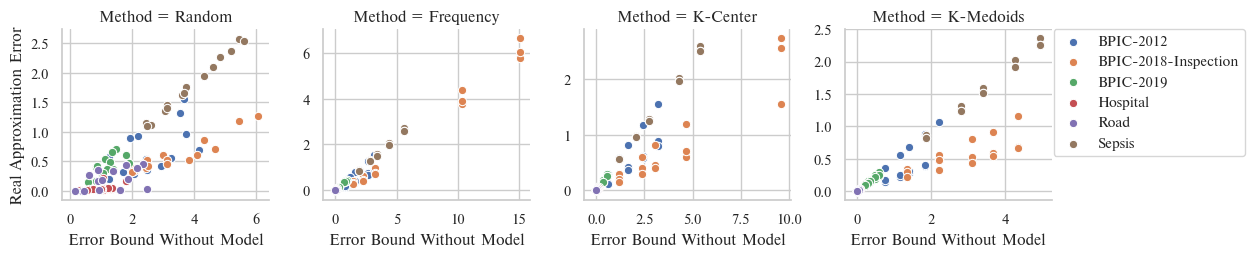}
    \caption{ Scatter plots of the estimated maximum error bounds and the real approximation error using different proxy-set generation methods.}
    \label{fig:withoutModel}
\end{figure}

Given that there is no causal relation, we investigate, using the described proxy-set generation methods and event logs, the strength of the correlation between the estimated maximum error and the effective approximation error when using the proxy-set.
In \autoref{fig:withoutModel}, we show the scatter plots of these two values, for each method, using different colors for the different event logs.
%\begin{table*}[tb]
%	\centering
%\caption{New table to support \autoref{fig:withoutModel}
%			\label{tab:withoutModel}
%		}
%		\includegraphics[width=\textwidth] {AvgCor.JPG}
%\end{table*}
Moreover, the corresponding Pearson correlation coefficients are presented in \autoref{tab:withoutModel}.
\begin{wraptable}{r}{5.75cm}
\vspace{-15pt}
\centering
\caption{Pearson correlation coefficients between the estimated bounds and the real approximation errors for different methods and event logs.  
\label{tab:withoutModel}
}
\vspace{-3pt}
	\includegraphics[width=0.435\textwidth] {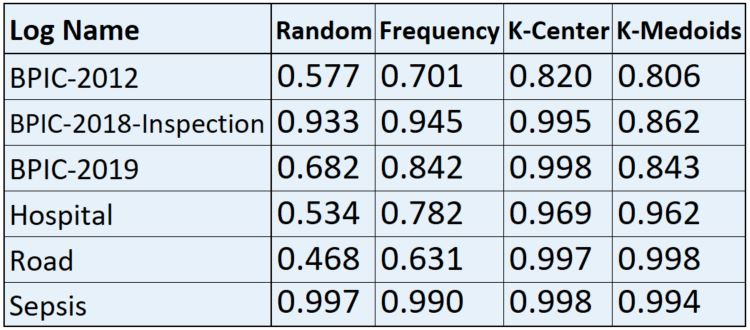}
\vspace{-10pt}
\end{wraptable}

Interestingly, for frequency-based sampling, K-Center, and K-Medoids, we observe a strong correlation between the estimated maximal approximation error and the effective approximation error.
For random sampling, as expected, the correlation is less strong, particularly for the Hospital-Billing and Road logs.
For all event logs, the highest correlation is achieved by the K-Center method.

\begin{figure}[tb]
	\centering
	\includegraphics[width=\textwidth]{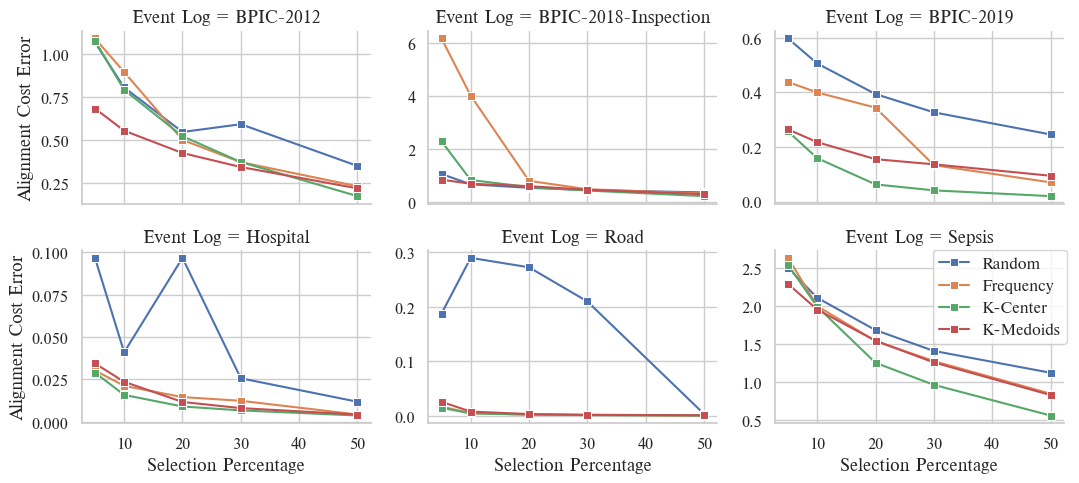}
	\caption{Effect of increasing the selected percentage of variants on approximated alignments' accuracy for different methods.
		\label{fig:Accuracy}
	}
\end{figure}

In \autoref{fig:Accuracy}, we show the effect of choosing different proxy-set generation methods and different percentages of variants in the event log on the accuracy of approximated alignment costs. 
As we expect, the K-Center and K-Medoids approaches provide the highest accuracy.
Therefore, using these approaches, we are able to generate more suitable proxy-sets and consequently obtain more accurate approximations. 
Moreover, results indicate that the alignment cost error is reduced by increasing the proxy-sets' size (i.e., the selection percentage). 
However, for some event logs, specifically if the variants in the event log are similar, this reduction is not always significant, i.e., by just using a few  trace-variants  we already obtain an accurate approximation. 
Thus, the approximation's provided bounds help users adjust the setting more efficiently, as a user may make the choice of increasing the proxy set size, thus increasing the accuracy of approximation.

%\begin{table}[tb]
%	\centering
%\caption{Required of different proxy-set selection methods.   
%			\label{tab:timeDist}
%		}
%		\includegraphics[width=0.925\textwidth] {TimeDistributation.JPG}
%\end{table}

\subsubsection{Conformance Checking Performance Improvement}
\begin{figure}[tb]
    \centering
    \begin{subfigure}[b]{\textwidth}\centering
    \includegraphics[width=0.95\textwidth]{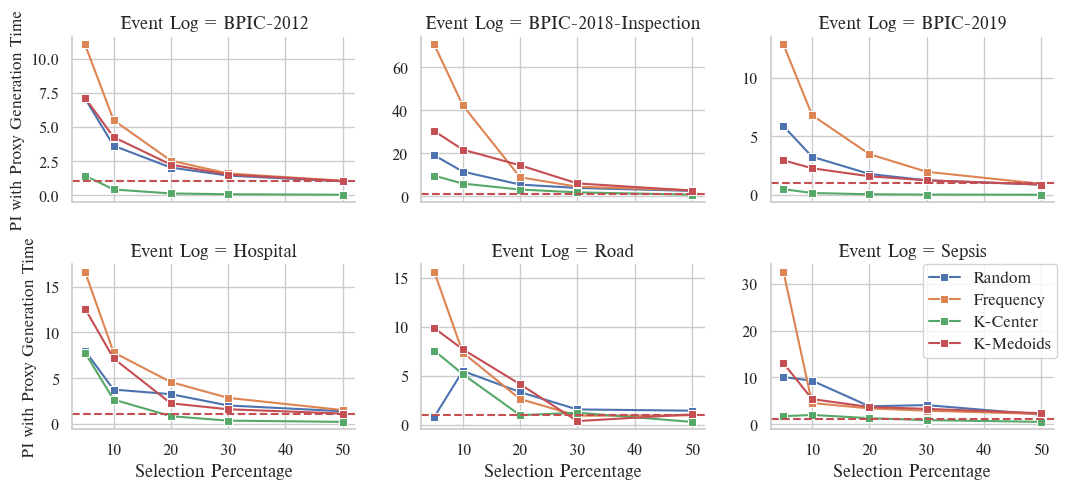}
    \caption{Performance improvement with consideration of proxy-set generation time.}
    \label{fig:PIwhith}
    \end{subfigure}
	\begin{subfigure}[b]{\textwidth}\centering
    \includegraphics[width=0.95\textwidth]{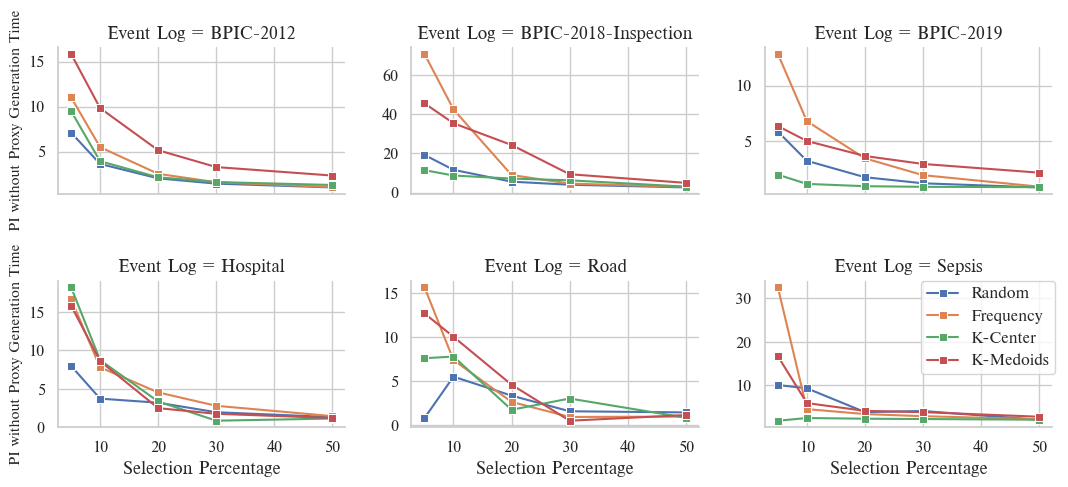}
    \caption{Performance improvement without consideration of proxy selection time.}
    \label{fig:PIwhithout}
    \end{subfigure}
    \caption{Effect of increasing the selected percentage of variants on performance improvement of different proxy selection methods.}
\end{figure}

Here, we analyze the time performance of different proxy-set generation methods.
Figures \autoref{fig:PIwhith} and \autoref{fig:PIwhithout} show the conformance checking performance improvement using the proposed approach.
To compute the \emph{performance improvement} $PI$, we divide conventional alignment computation time by the alignment approximation time (both including and excluding proxy-set generation time).
% \begin{equation}\scriptsize
%     PI= \frac{\text{Normal Alignment Time}}{\text{Approximated Alignment Time}\; (+ \text{Proxy-set Generation Time})}
% \end{equation}
A higher $PI$-value indicates a higher performance improvement and a $PI$-value less than $1$ indicates that there is no improvement. 
The greatest improvement (when we consider the proxy generation time) is achieved by using the frequency-based method as it quickly selects variants and generates the proxies.
The Random method has a lower $PI$-value because it may select some variants that usually need more time to compute their alignments. 
Furthermore, the results indicate that by increasing the percentage of the size of proxies, the performance improvement is reduced.
In some cases, we do not improve the performance when the proxy generation time is considered. 
Thus, it is crucial to avoid selecting too many traces as a proxy.
Generally, the proxy generation time for K-Center and K-Medoids methods is high, especially when the z
size of proxy is high. 
But, if we separate the proxy generation time (that is possible in some applications as explained in \autoref{sec:intro}), we improve the conformance checking procedure's performance.
%More information about the required time for different phases of approximation is presented in \autoref{tab:timeDist}. 

 \subsubsection{Efficiency of the Proposed Lower Bound}
 Finally, in the last experiment, we analyze the role of the lower bound without $\modelSubset$, i.e., $\max(0,\min\limits_{\sequence'{\in}\lang(\model)}(|\sequence'|){-}|\sequence|){+}|\sequence_{\downarrow_{\univAct{\setminus}\univAct_{\model}}}|$ and the lower bound that uses $\modelSubset$, i.,e., $\max\limits_{\sequence'{\in}\proxySet}(\optimalAlignmentCost(\sequence',\lang(\model)){-}\editDistance(\sequence,\sequence'))$) in computing the final lower bound. 
\autoref{tab:lb} shows the percentage of traces that have higher values using the different bounds.
In case the highest value is returned by both methods, we consider both of them as the used bound.
\begin{wraptable}{r}{4.25cm}
	\centering
	\vspace{0pt}
\caption{ Average percentage of times that lower bounds have the highest value.
\label{tab:lb}
		} \vspace{-5pt}
		\includegraphics[width=0.28\textwidth] {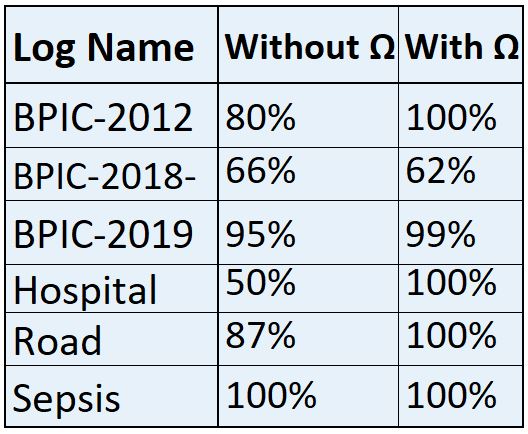}
		\vspace{-50pt}
\end{wraptable}
The results show that in most cases, it is sufficient to use the lower bound, which is based on the proxy-set and its alignments. Consequently, using the new proposed method for bound computation, we will have tighter bounds.

\section{Conclusion}
\label{sec:conclusion}

In this paper, we propose to select a set of traces, compute their alignments and use these to approximate the alignments of other traces in the event log. 
Furthermore, we have shown that we can derive bounds for the so-introduced approximation error, independently of the model. 
Additionally, we prove that based on the selected traces (i.e., a proxy-set), we can provide a bound for the approximation error that helps users estimate the approximation error and thus aids in selecting an appropriate proxy set. 
The experiments on the real event logs indicate, that by using the proposed instance selection methods, we are able to reduce the maximum and the average error in the alignment cost approximation. 
Besides, by increasing the number of the selected traces, the average possible error is reduced. But, for certain event logs, this reduction is not significant, which shows we are able to select a few traces and have an accurate approximation. 

We used the approximation solutions for K-Center and K-Medoids problems in this work. 
However, it is useful for some applications to compute optimal solutions and find the best K variants. 
Moreover, it is beneficial to provide an incremental approach that keeps selecting the traces until we guarantee a tight bound for the approximation error.

\bibliography{bibliography}

\end{document}